\documentclass[10pt]{article}
\usepackage{graphicx}
\usepackage[
  backend=biber,
  style=numeric,
  sorting=none,
  maxnames=6,   
  minnames=1    
]{biblatex}
\usepackage{booktabs,tabularx,makecell}
\usepackage{indentfirst,csquotes}
\usepackage{amssymb,amsthm,amsmath}
\usepackage{xcolor,paralist,hyperref,titlesec,fancyhdr,etoolbox}
\usepackage{lipsum}

\title{Seasonal and periodic patterns of ischemic heart disease in New York using the Variable Multiple Bandpass Periodic Block Bootstrap}

\author{
  Yineng Chen\thanks{Corresponding author.}\\
  \texttt{ychen77@albany.edu}
  \and
  Edward Valachovic \\
  \texttt{evalachovic@albany.edu}
}

\date{Department of Epidemiology and Biostatistics, College of Integrated Health Sciences,\\
University at Albany, State University of New York, One University Place, Rensselaer, NY 12144}

\begin{document}
\maketitle

\begin{abstract}
Seasonal patterns of the incidence, hospital visits, and mortality of ischemic heart disease (IHD) have been widely reported. This study aims to investigate seasonal and periodic patterns of IHD hospitalizations in New York using a novel bootstrap approach, the Variable Bandpass Periodic Block Bootstrap (VBPBB) method. Using a bandpass filter, VBPBB isolates the periodically correlated (PC) component of interest from other PC components and noise before bootstrapping, preserving correlation structures and yielding more precise 95\% confidence intervals than existing periodic bootstrapping methods. We examine weekly, monthly, and annual patterns, along with their harmonic frequencies, in the IHD hospitalization. In addition to the pre-defined frequencies, we also examine the frequencies with the highest amplitudes in the periodogram. By aggregating bootstrap results from statistically significant PC components, a 95\% CI band that preserves multiple periodic correlation structures was obtained. Statistically significant variation was observed for the weekly, annual component, and its 2nd, 3rd, 5th, and 6th harmonics. CI bands obtained from VBPBB were much narrower than those from existing periodic bootstrapping methods. VBPBB substantially improves the precision of periodic mean estimates while preserving periodic correlation structures, making it suitable for time series with multiple periodic patterns and high noise, such as in environmental or healthcare data. 
\end{abstract} 

\section*{Introduction}
Ischemic heart disease (IHD) remains the leading cause of death worldwide and a significant contributor to global disability \cite{rothGlobalBurdenCardiovascular2020a}. While prior studies have documented seasonal patterns in IHD, many rely on pre-specified seasonal categories or simple parametric models that may overlook complex, overlapping, or nonstandard periodic structures \cite{manfrediniSeasonalWeeklyPatterns2009,shethIncreasedWinterMortality1999,abrignaniInfluenceClimaticVariables2009,reaveyExploringPeriodicityCardiovascular2013,lashariVariationAdmissionRates2015}. Understanding the Seasonal and periodic patterns of IHD is critical for effective resource allocation, preventive interventions, and public health planning. This study addresses these gaps by applying a novel approach—the Variable Multiple Bandpass Periodic Block Bootstrap (VMBPBB)—to isolate, test, and quantify periodic components in IHD hospitalizations in New York State from 2002 to 2023.

Bootstrapping is a statistical method for estimating the distribution of an estimator by resampling with replacement, leading to independent samples of the same length as the original dataset \cite{efronBootstrapMethodsAnother1979}. It is a non-parametric approach, particularly useful when the underlying population distribution is unknown or complex. However, in the presence of temporal correlation, such as in time series data, independently resampling individual observations can destroy correlation structures, leading to information loss. To address this, specialized techniques like block bootstraps have been developed, which resample data blocks instead of individual points.  A common approach is the Moving Block Bootstrap, introduced by Kunsch \cite{kunschJackknifeBootstrapGeneral1989}. While this method preserves the correlation within block size, it fails to maintain correlations between data points that are farther apart than the block length. Periodic components with period p show strong correlations between observations separated by $k \cdot p$ time points, where k is an integer. Such components are considered periodically correlated (PC). To accommodate such structure, the Seasonal Block Bootstrap proposed by Politis \cite{politisResamplingTimeSeries2001} constrains the block size to be a multiple of the period p, ensuring that all resampled blocks start at a consistent phase of the cycle. The Generalized Seasonal Block Bootstrap (GSBB) by Dudek et al. extends this idea and similarly aims to preserve periodic correlation in PC time series \cite{dudekGENERALIZEDBLOCKBOOTSTRAP2014}. Block bootstrap methods using such a strategy could be referred to as periodic block bootstrap (PBB) methods. However, due to noise and random error, the bootstrapped confidence intervals from PBB could be excessively wide, an issue that has been confirmed by previous simulations \cite{valachovicPeriodicallyCorrelatedTime2024}. 

The Variable Bandpass Periodic Block Bootstrap (VBPBB) extends PBB by treating time series as signals composed of additive periodically correlated (PC) components \cite{valachovicPeriodicallyCorrelatedTime2024}. Each PC component is associated with a specific frequency of interest and its harmonics. Harmonic frequencies are integer multiples of a fundamental frequency. For example, if a signal has a fundamental frequency of $f$, then its harmonics occur at $k \cdot f$, where k is an integer, and it is called the $k^{th}$ harmonics of the fundamental frequency. In practice, a PC component may deviate from a perfect sinusoidal wave confined to a single dominant frequency, often exhibiting variations at its harmonic frequencies. Therefore, this study examines both fundamental frequencies and their harmonics. Using a bandpass filter on the PC component frequencies, VBPBB filters the original time series into distinct PC components centered around predefined frequencies. Each component is then bootstrapped independently using a PBB approach, allowing for the preservation of periodic correlation and the generation of robust confidence intervals around periodic means. This method provides a non-parametric way to assess the statistical significance of periodicities without assuming linearity or specific model forms. Furthermore, many time series contain multiple periodically correlated (MPC) components. For example, a fundamental frequency and its harmonics often coexist within a signal.  Another example is hourly temperature data, which exhibits both daily periodic patterns driven by the Earth's rotation and seasonal patterns resulting from the Earth's orbit around the sun. Variable Multiple Bandpass Periodic Block Bootstrap (VMBPBB), an extension of VBPBB, forms a bootstrap of the MPC time series by aggregating the PC component bootstraps and can successfully preserve all periodic correlations within an MPC time series \cite{Valachovic_2025}. 

This study aims to investigate seasonal and periodic patterns of ischemic heart disease (IHD) hospitalizations in New York using a novel bootstrap approach. We apply the VBPBB to isolate and evaluate a series of periodically correlated (PC) components and subsequently use VMBPBB to aggregate the bootstrapped PC components, providing interval estimates for the overall IHD time series. This approach enables more accurate estimation and detection of significant periodic components, overcoming limitations of existing block bootstrapping methods for MPC time series.

\section*{Method}
\subsection*{Data Sources and Analysis}
Hospital admission data were obtained from the New York State (NYS) Department of Health’s Statewide Planning and Research Cooperative System (SPARCS) \cite{newyorkstatedepartmentofhealthStatewidePlanningResearch}, which captures approximately 95\% of all hospital discharges in NYS. The SPARCS database includes deidentified administrative information for each patient admission, including the date of admission and the principal diagnosis coded using the International Classification of Diseases (ICD). For this study, we retained hospitalizations with a principal diagnosis of ischemic heart disease (IHD), defined as ICD-9 codes 410–414 or ICD-10 codes I20–I25, occurring between January 1, 2002, and December 31, 2023. Hospitalization rate was calculated by dividing the number of hospital admissions by the total population and expressed as rates per 100,000 population. To address the presence of a long-term decreasing trend in the original time series, we fitted a linear regression to the data and subtracted the resulting trend component. This detrending step isolates short-term fluctuations while preserving periodic structure for spectral analysis. To identify frequencies of interest, we generated a periodogram from the detrended data, representing the time series in the frequency domain through its spectral density \cite{weiw.TimeSeriesAnalysis1989}. Frequency was selected based on prominent amplitudes in the periodogram and their alignment with known human activity patterns. PC components investigated in this study analyzed in this study correspond to annual, monthly, and weekly cycles. Harmonics of these frequencies (e.g., 2/365, 3/365) were also accessed. All analysis was conducted using R (version 4.4.1).

\subsection*{Variable Multiple Bandpass Periodic Block Bootstrap}

To generate 95\% confidence interval bands, we applied the Variable Bandpass Periodic Block Bootstrap (VBPBB) and its extension, the Variable Multiple Bandpass Periodic Block Bootstrap (VMBPBB) method, a block bootstrap method for time series with multiple periodically correlated (MPC) components. Unlike existing periodic bootstrap methods that resample the original time series, including PC components of interest, other components, and noise that leads to increased bootstrap variability, VMBPBB uses bandpass filters to separate each PC component from noise and other interference and then resamples each filtered PC component respectively. Upon aggregating the PC component bootstraps, a bootstrap for the MPC time series is formed, preserving correlation structures of all periodic components and reducing interference from noise.

The VMBPBB approach uses the Kolmogorov-Zurbenko (KZ) filter and its extensions to separate each PC component of interest from the noise \cite{Valachovic_2025}. The Kolmogorov-Zurbenko Fourier Transform (KZFT) filter, an extension of the KZ filter. was applied in this study. The KZFT filter is a bandpass filter that passes a particular range of frequencies around the corresponding frequency of the PC component and rejects, attenuates, or suppresses other frequency components from the input signal. KZFT filters have been widely applied across various domains, including ozone concentration (Tsakiri and Zurbenko), air quality (Kang et al.), global temperature (Lou and Zurbenko), atmospheric phenomena (Zurbenko and Potrzeba), and climate studies (Zurbenko and Cyr) \cite{kangApplicationKolmogorovZurbenko2013,tsakiriDeterminingMainAtmospheric2010,zurbenkoRestorationTimeSpatialScales2012,zurbenkoTidalWavesAtmosphere2010,zurbenkoClimateFluctuationsTime2011}. In the field of public health, notable applications include research on diabetes (Arndorfer and Zurbenko), skin cancer (Valachovic and Zurbenko), and COVID-19 (Valachovic and Shishova) \cite{arndorferTimeSeriesAnalysis2017,valachovicSkinCancerSolar2013,valachovicSeasonalPeriodicPatterns2025}.  

The KZ filter is an iterated moving average (MA) filter defined by two arguments: $m$, a positive integer representing the window length, and $k$, the number of iterations applied \cite{i.g.zurbenkoSpectralAnalysisTime1986}. As a low-pass filter, the KZ filter effectively attenuates signals with frequencies at or above $\frac{1}{m}$, while preserving lower-frequency components. The KZFT filter is derived from the KZ filter by applying it to the Fourier transform \cite{yangKolmogorovZurbenkoFilters2010a}. It is a bandpass filter with three arguments: $m$ (window length), $k$ (number of iterations), and $v$ (the center frequency around which the filter is applied). A KZFT filter applied to a random process $\{X(t):t\in T \}$ is demonstrated in the following equation:

$$KZFT_{(m,k,v)}= \sum_{u=-\frac{k(m-1)}{2}}^{\frac{k(m-1)}{2}} \frac{a_u^{m,k}}{m^k}  e^{-i2mvu} X(t+u)$$

where the coefficients $a_u^{m,k}$ are the polynomial coefficients from:

\[
\sum_{r=0}^{k(m+1)} z^r \, a_{\,r - \tfrac{k(m-1)}{2}}^{m,k}
= (1+z+\cdots+z^{m-1})^k
\]

The KZFT is a symmetric band pass filter around frequency $v$ and its energy transfer function at a frequency $\lambda$ is shown below:

\[
\lvert B(\lambda - v) \rvert^2 
= \left( \frac{\sin\!\big(\pi m(\lambda - v)\big)}{m \, \sin\!\big(\pi(\lambda - v)\big)} \right)^{2k}
\]

KZFT filters can be implemented using the KZFT function in the KZA package \cite{briancloseigorzurbenkomingzengsunKzaKolmogorovZurbenkoAdaptive2020}. 

In this study, we investigated annual, monthly, and weekly PC components, along with their corresponding harmonics. In addition to these predefined cycles, we identified and analyzed the ten frequencies with the highest amplitudes in the periodogram, excluding those already mentioned. KZFT filters were then applied separately to each PC component of interest derived from the IHD hospitalization time series, with the centering frequency argument ($v$) set to match the frequency of each targeted component. 

To isolate a single PC component in each KZFT bandpass filter, the filter window length ($m$) must be set to avoid overlap with adjacent frequencies to be rejected. Previous simulation studies have shown that increasing $m$ narrows the filter’s bandwidth, effectively suppressing frequencies outside a narrow range centered around the target frequency $v$ \cite{valachovicPeriodicallyCorrelatedTime2024}. A moving average filter with window size m replaces each value with the average of $m$ surrounding points, requiring sufficient data to compute each smoothed value. Consequently, the first and last $m/2$ observations are lost, and to ensure feasibility, $m$ must not exceed a length that is a significant portion of the total time series length \cite{valachovicPeriodicallyCorrelatedTime2024,Valachovic_2025}. Specifically, for any two adjacent PC components of interest with periods $p_1$ and $p_2$ , and corresponding frequencies $v_1=1/p_1$  and $v_2=1/p_2$, the initial value for the filter window length m is set as the smallest odd integer greater than $m^*= \frac{4}{|v_1-v_2 |} = \frac{4(p_1 p_2)}{|p_1-p_2 |}$.  Periodograms were generated for each filtered PC component to evaluate the effectiveness of the KZFT filtering. If noticeable amplitude remains at frequencies other than the target frequency $v$, the window length $m$ is increased to the nearest odd integer greater than $1.5 \times m^*$. Since each iteration of the KZFT filter effectively increases the data requirement, the number of iterations ($k$) is fixed at 2 for all filters in this study. 

After separating the original time series into multiple PC components using bandpass filters, the VBPBB method performs a periodic bootstrap on each PC component independently. For a given period $p$, the VBPBB method partitions the series of $n$ observations into $p$ non-overlapping subsets corresponding to periodic positions, then independently resamples from each subset with replacement until $n$ observations are selected — preserving intra-period correlation. This process is repeated across $B=1000$ bootstrap replicates. The periodic mean at each time point is computed for each replicate to construct a 95\% confidence interval band. PC components whose 95\% confidence intervals band cannot fit a horizontal line were considered statistically significant. In such a case, the 95\% confidence interval band would reject a hypothesis of a periodic correlation with zero amplitude at this frequency. Numerically, this occurs when the minimum value of the upper confidence interval band is smaller than the maximum value of the lower confidence interval band. This could be referred to as the “horizontal-line” significance criterion. The $B$ bootstraps of each PC component with statistically significant 95\% confidence intervals are subsequently recombined to form $B$ bootstraps of the MPC series and then construct a 95\% confidence interval band for the combined significant PC components.

\subsection*{Existing PC bootstrapping approaches}

As a comparison to the VBPBB method, we implemented a simplified version of the Generalized Seasonal Block Bootstrap (GSBB) to generate periodic resamples of the time series \cite{dudekGENERALIZEDBLOCKBOOTSTRAP2014}. For a given period (e.g., 7 for weekly cycles), the original series was divided into disjoint strata: values occurring every p time points were grouped so that each stratum contained observations from the same phase of the cycle (e.g., all Mondays, all Tuesdays, etc.). Within each stratum, values were independently resampled with replacement to preserve the periodic structure. This resampling was repeated $B=1000$ times to produce a bootstrap series of the same length as the original. The 95\% confidence interval band was constructed by computing the periodic mean at each time point across all bootstrap replicates.

\section*{Results}

\begin{figure}[ht] 
    \centering
    \includegraphics[width=0.7\textwidth]{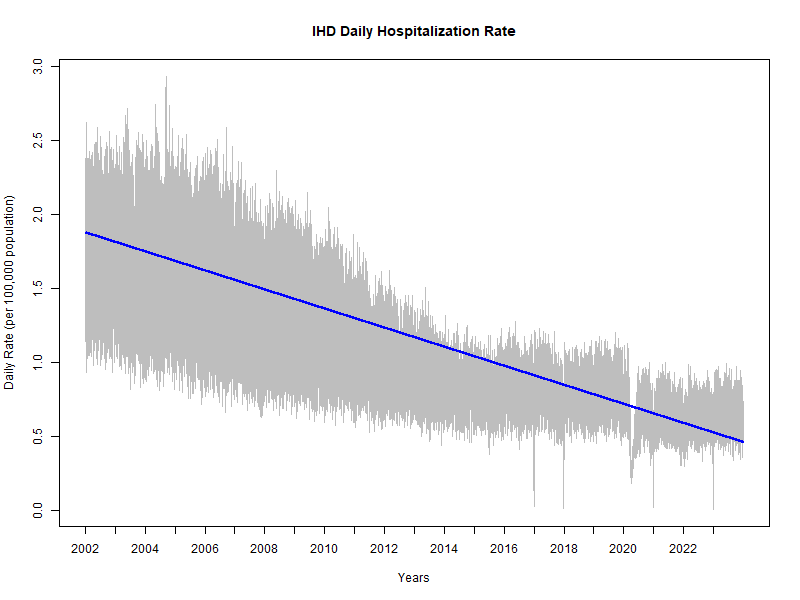}
    \caption{Daily Rate of IHD Hospitalization (per 100,000 population) of New York State.}
    \label{fig:Original}
\end{figure}

The daily ischemic heart disease (IHD) hospitalization rates in New York State from 2002 to 2023 displayed a clear decreasing trend, with notable disruptions during the COVID-19 pandemic (Figure~\ref{fig:Original}). In total, 8,035 days of data were analyzed, with a mean hospitalization rate of 1.17 (SD = 0.55) per 100,000 population. 

\begin{figure}[ht]
    \centering
    \includegraphics[width=0.7\textwidth]{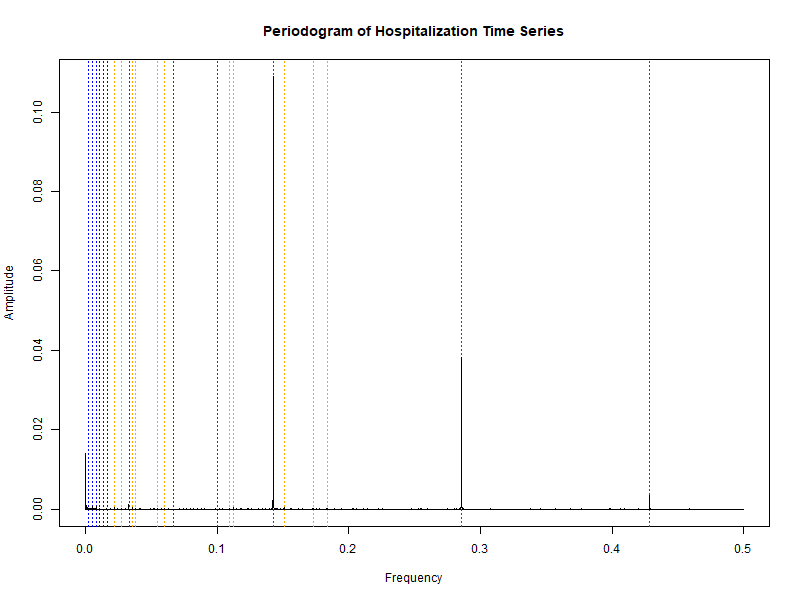}
    \caption{Periodogram of frequencies present in the IHD hospitalization rate time series.}
    \label{fig:Periodogram_all}

    \vspace{0.5em}
    \footnotesize\emph{Note:} Red dotted lines: weekly frequency and harmonics; green dotted lines: monthly frequency and harmonics; blue dotted lines: yearly frequency and harmonics; yellow dotted lines: top ten highest-amplitude frequencies in the periodogram (excluding those above)
\end{figure}

\begin{figure}[ht]
    \centering
    \includegraphics[width=0.7\textwidth]{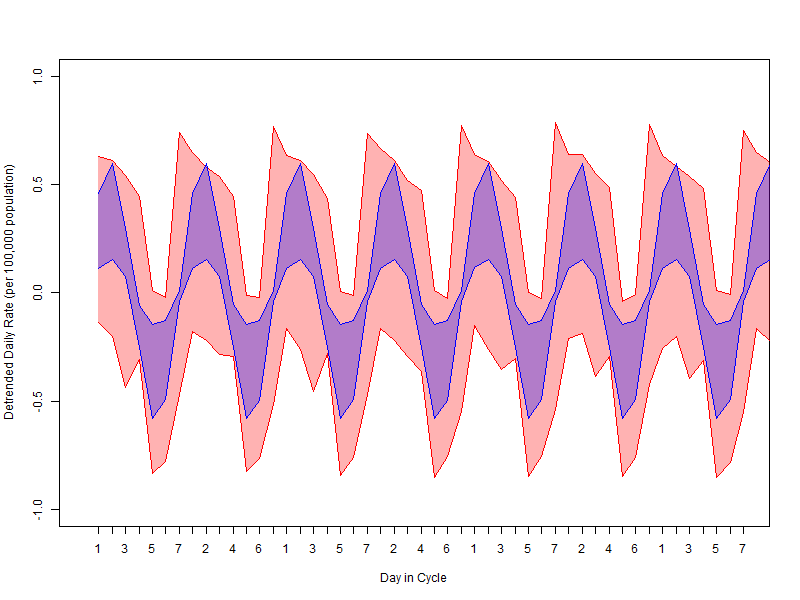}
    \caption{PBB (red) and VBPBB (blue) 95\% CI bands for the hospitalization rate weekly mean variation. }
    \label{fig:Week_1_band}
\end{figure}

\begin{figure}[ht]
    \centering
    \includegraphics[width=0.7\textwidth]{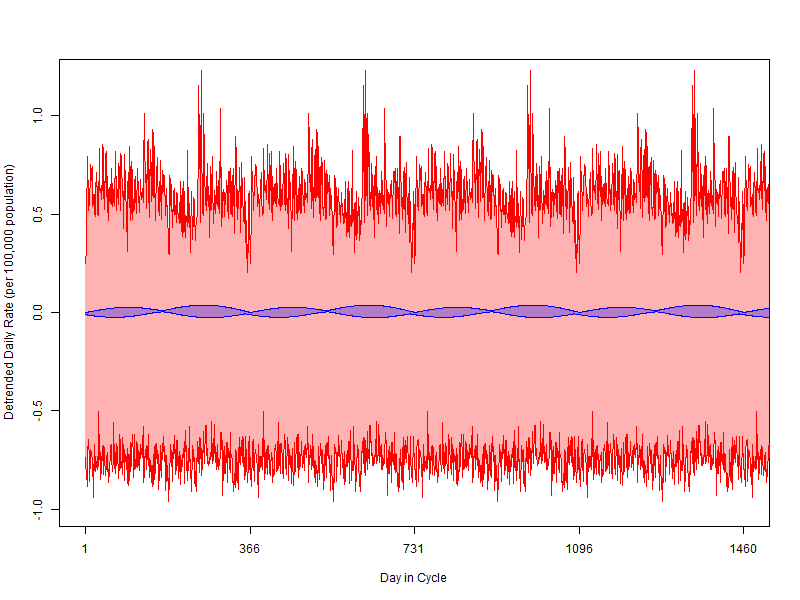}
    \caption{PBB (red) and VBPBB (blue) 95\% CI bands for the annual hospitalization rate mean variation.}
    \label{fig:Annual_1_band}
\end{figure}

\begin{figure}[ht]
    \centering
    \includegraphics[width=0.7\textwidth]{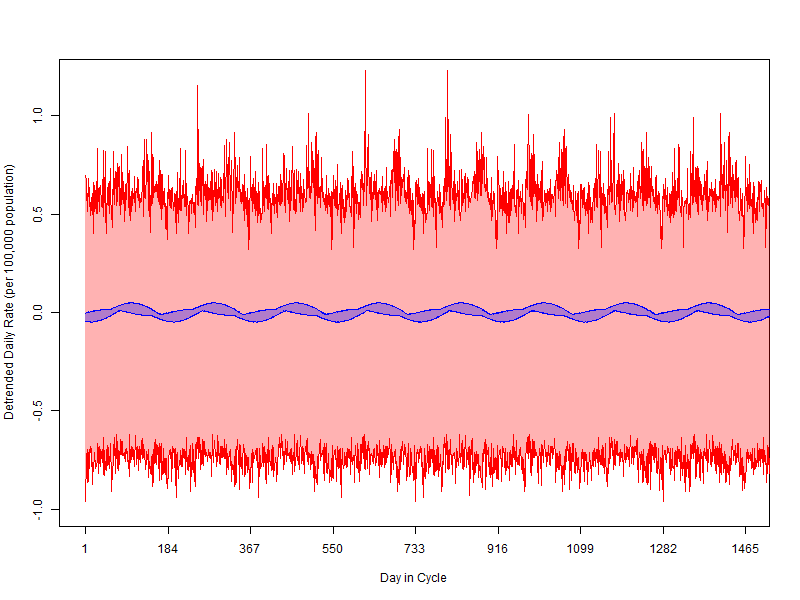}
    \caption{PBB (red) and VBPBB (blue) 95\% CI bands for the annual second harmonic mean variation.}
    \label{fig:Annual_2_band}
\end{figure}

\begin{figure}[ht]
    \centering
    \includegraphics[width=0.7\textwidth]{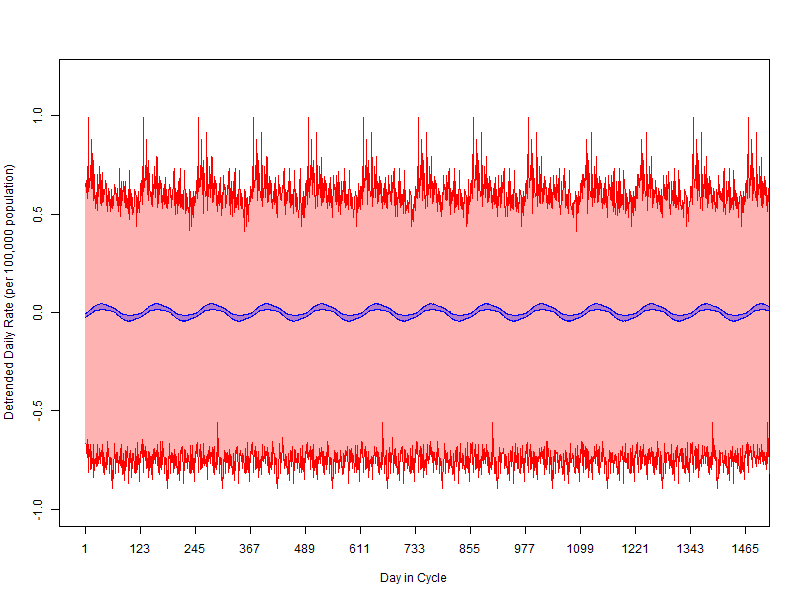}
    \caption{PBB (red) and VBPBB (blue) 95\% CI bands for the annual third harmonic mean variation.}
    \label{fig:Annual_3_band}
\end{figure}

\begin{figure}[ht]
    \centering
    \includegraphics[width=0.7\textwidth]{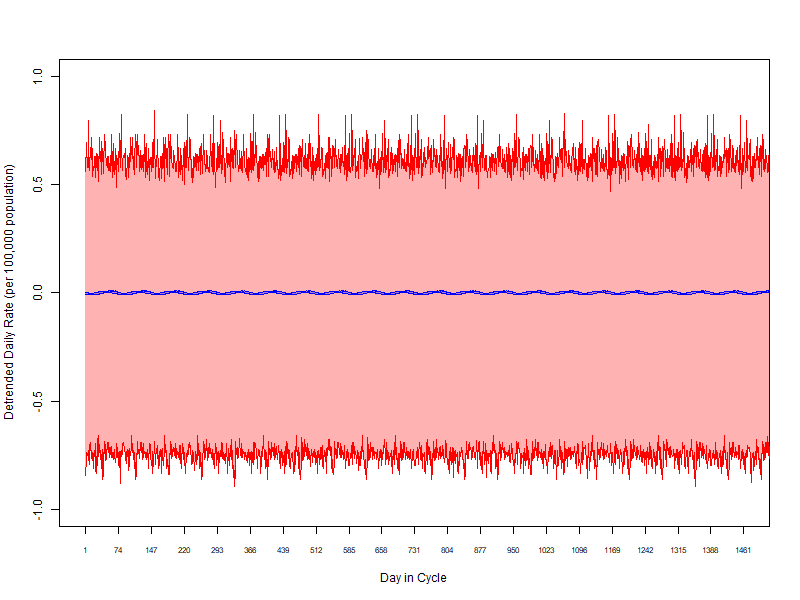}
    \caption{PBB (red) and VBPBB (blue) 95\% CI bands for the annual fifth harmonic mean variation.}
    \label{fig:Annual_5_band}
\end{figure}

\begin{figure}[ht]
    \centering
    \includegraphics[width=0.7\textwidth]{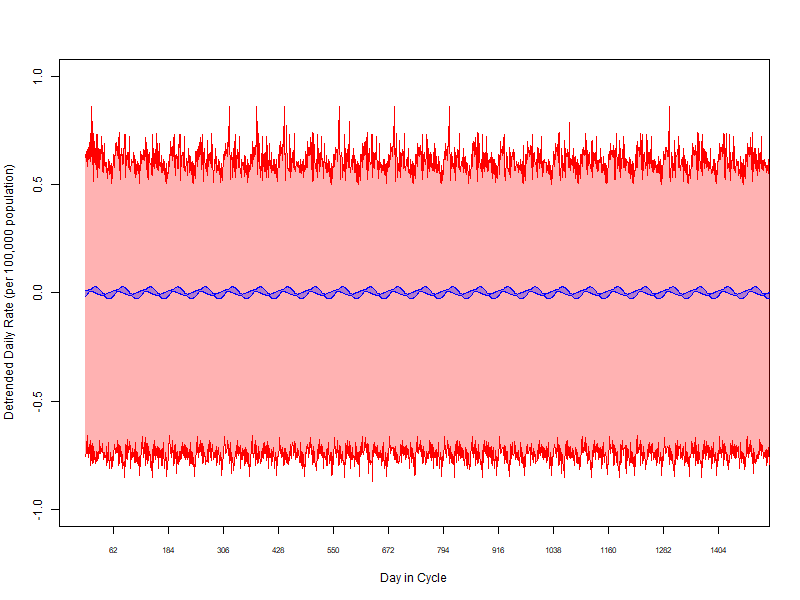}
    \caption{PBB (red) and VBPBB (blue) 95\% CI bands for the annual sixth harmonic mean variation.}
    \label{fig:Annual_6_band}
\end{figure}

\begin{figure}[ht]
    \centering
    \includegraphics[width=0.7\textwidth]{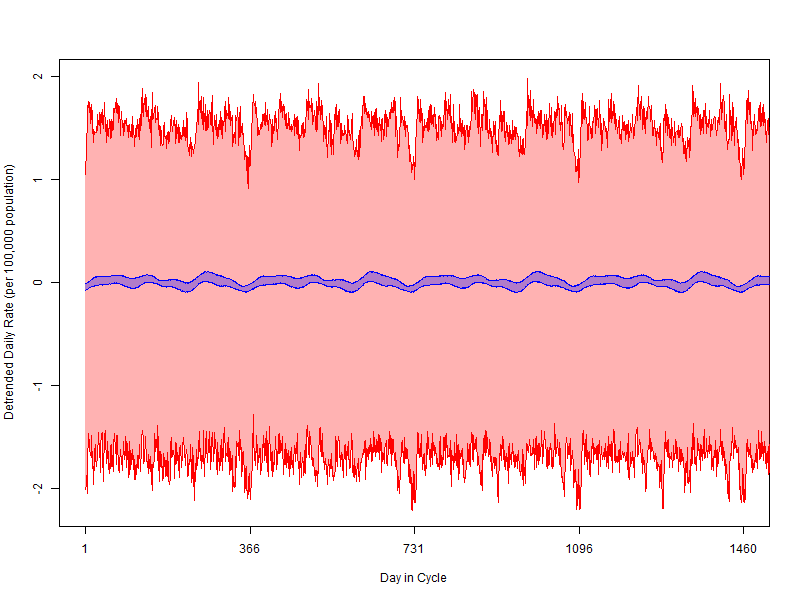}
    \caption{VBPBB 95\% CI band for the hospitalization rate periodic mean, combining annual and all significant harmonic components (blue), versus PBB 95\% CI band for the annual mean variation only (red).}
    \label{fig:MVBPBB}
\end{figure}

Figure~\ref{fig:Periodogram_all} presents the periodogram of the detrended IHD hospitalization series. The periodogram revealed distinct peaks at weekly frequency and its harmonic frequencies. Peaks at annual frequency and its harmonic frequencies were also observed (Supplementary Figure S1 – S3).  The top ten additional high-amplitude frequencies besides pre-defined frequencies were also denoted in the periodogram. Table~\ref{tab:kzft_vbpbb} summarizes the results and KZFT arguments for the VBPBB and GSBB bootstrap analyses of the IHD hospitalization rate time series. The weekly component (p = 7 days) showed a VBPBB 95\% CI crest range of 0.156 to 0.597 and trough range of -0.582 to -0.150, compared to GSBB ranges of -0.100 to 0.649 and -0.762 to -0.056. The CI band width ratio = 3.0 comparing GSBB to VBPBB, meaning the size of the CI band produced by GSBB is approximately three times the size of that from VBPBB. Statistically significant variation was also observed for the annual component (p = 365 days; crest: 0.003 to 0.009, trough: –0.011 to 0.000; CI band width ratio = 35.6) and its 2nd, 3rd, 5th, and 6th harmonics (CI band width ratios = 32.0, 55.5, 179.1, and 66.7, respectively). Across all components, the VBPBB band (blue) is noticeably narrower than the PBB band (red), indicating improved precision. The weekly component peaks on Wednesdays of the week and reaches its lowest value on Sundays (Figure~\ref{fig:Week_1_band}). The annual component peaks around July 20 and reaches a trough around December 26. The second, third, fifth and sixth annual harmonics show shorter cycle lengths, with multiple peaks occurring within a single year (Figures~\ref{fig:Annual_1_band}-\ref{fig:Annual_6_band}). No statistically significant variation was observed among the top-amplitude frequencies other than the pre-defined frequencies. A summary of the KZFT arguments and the 95\% CI were shown in Supplementary Table 1S. Aggregating annual PC components with all their significant harmonics using the MVBPBB method yielded an overall 95\% CI band for the IHD periodic mean that captured the combined annual and harmonic variation (Figure~\ref{fig:MVBPBB}). Compared to the CI band obtained using PBB in red, the CI band from VMBPBB in blue was much narrower. The combined signal remains relatively stable throughout the year, with only modest but significant fluctuations and a small peak appearing toward the end of the annual cycle.

\section*{Discussion}

In this study, we applied VBPBB to isolate frequencies of interest to quantify periodic mean variation in IHD hospitalization rates in New York State and compared its performance with the existing PBB. While PBB was able to identify one significant PC component, the weekly component, VBPBB was able to identify the weekly, annual component, along with annual second, third, fifth, and sixth harmonics. Compared with PBB, the VBPBB approach yielded markedly narrower confidence intervals while preserving periodic correlation structures, indicating substantially improved precision in estimating periodic mean variation. This improvement is particularly notable given the presence of high noise, long-term trends, and disruptive events such as the COVID-19 pandemic. Several frequencies with high amplitudes were identified but ultimately found to be insignificant PC components, reflecting the high noise level in the study data. Previous studies have reported associations between cardiovascular diseases and natural disasters, power outages and air pollution \cite{dengIndependentSynergisticImpacts2022a,devitaImpactClimateChange2024,linImmediateEffectsWinter2021,krittanawongPM25CardiovascularDiseases2023,yangHospitalizationRisksAssociated2025}, which can interfere with the periodic pattern of IHD hospitalization rates. Another major disruption in the time series was the COVID-19 pandemic, resulting in a sharp decline in hospitalization rates in early 2020. By isolating periodic components from noise before resampling, VBPBB effectively reduces the influence of non-periodic anomalies, ensuring that large, irregular events occurring outside the target frequencies do not distort periodic mean estimates or their confidence interval bounds.

Another advantage of VBPBB is that it can apply to data with long-term trends. A long-term trend in the data can be viewed as part of a PC component with an extremely large period, corresponding to a very low frequency. When VBPBB isolates a PC component of interest (e.g., annual, weekly), it removes frequency components outside the target range, meaning the trend frequency is excluded from the resampling process and does not bias the estimation of the periodic mean. In theory, VBPBB could also be applied to estimate such ultra-low-frequency components (e.g., multi-year trends). However, isolating these frequencies requires extremely large values of the KZFT arguments m or k, which causes substantial data loss ($k \times \frac{m}{2}$ observations removed). This makes it impractical unless an exceptionally large dataset is available. In this study, we detrended the data primarily to improve visualization, making the confidence interval (CI) bands appear nearly horizontal. Small residual trend signals remained, as indicated by faint low-frequency peaks in the periodogram, but their frequencies were much lower than the smallest frequency investigated (1/365) and thus did not interfere with the periodic mean estimation. Given the limited time span of available data, we could not investigate longer periodicities, such as 5-year or 10-year cycles.

In fact, PBB can be seen as a special case of VBPBB, where the bandpass filter is so wide that it passes all frequencies ($m$=1, $k$=1). However, the improvement in estimation of VBPBB to PBB in this work demonstrates an advantage of applying a bandpass filtering prior to block bootstrapping. This result is consistent with the previous simulation study \cite{valachovicPeriodicallyCorrelatedTime2024} and previous application studies \cite{valachovicSeasonalPeriodicPatterns2025,maioTemporalModelingNitrogen2024}. VBPBB has several limitations in its design in comparison to other PBB methods. VBPBB performance is tied to the selection of arguments used for bandpass filtration, and results may differ depending on the choice of KZFT filter arguments. This led to the problems of how to find the optimal KZFT filter arguments and whether the improved performance has a limit. Our research team is working on these problems to further improve the VBPBB method. Consequently, the precision of this method is expected to improve as additional data becomes available. Nevertheless, the present results still clearly demonstrate the presence of annual and other periodic components in the IHD hospitalization rate and provide a more precise estimation by combining multiple PC components.

In conclusion, this study demonstrates that the VBPBB method offers substantially improved precision in estimating periodic mean variation of IHD hospitalization rates compared with PBB, highlighting the value of VBPBB for analyzing complex health time series where both seasonality and irregular disturbances coexist. This study could be a preliminary step in studies where periodicity is not the primary focus. By correctly identifying and removing periodic components, future research could reveal other important features of time series data, such as long-term trends, discontinuities, or interruptions caused by events like COVID-19, changes in clinical practice, or policy shifts. Future research with longer observation periods could enable investigation of very low-frequency trends and multi-year cycles. In addition, the development of a standardized approach for selecting optimal KZFT filter parameters would further enhance the performance and applicability of the VBPBB method.

\begin{table}[ht]
\centering
\caption{Summary of Target Frequencies, Periods, KZFT Settings, and Result}
\label{tab:kzft_vbpbb}

\resizebox{\textwidth}{!}{%
\begin{tabular}{l c c c c c c c c}
\hline
\multicolumn{1}{c}{} & \multicolumn{3}{c}{KZFT Arguments} & 
\multicolumn{2}{c}{VBPBB 95\% CI Band} & 
\multicolumn{2}{c}{GSBB 95\% CI Band} & 
\multicolumn{1}{c}{Ratio} \\
\cline{2-4} \cline{5-6} \cline{7-8}
Component & $v$ & $m$ & $k$ & Crest & Trough & Crest & Trough & GSBB/VBPBB \\
\hline
Weekly                & 1/7   & 487  & 2 & (0.156, 0.597)* & (-0.582, -0.15)* & (-0.1, 0.649)  & (-0.762, -0.056)*   & 3.0 \\
2nd weekly harmonic   & 2/7   & 165  & 2 & (-0.314, 0.324) & (-0.35, 0.314)   & (-0.698, 0.629) & (-0.776, 0.546)    & 2.0 \\
3rd weekly harmonic   & 3/7   & 203  & 2 & (-0.11, 0.126)  & (-0.12, 0.107)   & (-0.693, 0.632) & (-0.749, 0.532)    & 5.6 \\
Monthly               & 1/30  & 1771 & 2 & (-0.002, 0.007) & (-0.009, 0.002)  & (-0.661, 0.581) & (-0.713, 0.517)    & 52.1 \\
2nd monthly harmonic  & 2/30  & 1077 & 2 & (-0.001,0.013)  & (-0.013,0)       & (-0.681,0.581)  & (-0.729,0.525)     & 81.5 \\
3rd monthly harmonic  & 3/30  & 631  & 2 & (-0.007,0.009)  & (-0.009,0.007)   & (-0.686,0.579)  & (-0.747,0.534)     & 72.2 \\
Annual                & 1/365 & 2921 & 2 & (0.003,0.009)*  & (-0.011,0)*      & (-0.502,0.504)  & (-0.820,0.205)     & 35.6 \\
2nd annual harmonic   & 2/365 & 1461 & 2 & (0.009,0.037)*  & (-0.036,-0.009)* & (-0.614,0.714)  & (-0.740,0.319)     & 32.0 \\
3rd annual harmonic   & 3/365 & 1461 & 2 & (0.014,0.044)*  & (-0.044,-0.014)* & (-0.558,0.513)  & (-0.771,0.404)     & 55.5 \\
4th annual harmonic   & 4/365 & 1461 & 2 & (-0.002,0.019)  & (-0.018,0.002)   & (-0.634,0.641)  & (-0.740,0.468)     & 51.1 \\
5th annual harmonic   & 5/365 & 2191 & 2 & (0.003,0.007)*  & (-0.007,-0.003)* & (-0.631,0.573)  & (-0.734,0.467)     & 179.1 \\
6th annual harmonic   & 6/365 & 1461 & 2 & (0.006,0.019)*  & (-0.018,-0.006)* & (-0.652,0.581)  & (-0.759,0.483)     & 66.7 \\
\hline
\end{tabular}
} 
\footnotesize{Note:} Indicates PC components that meet the “horizontal-line” significance criterion, (minimum of upper 95\% CI band $<$ maximum of lower 95\% CI band), indicating temporal variation in the component. 
\end{table}

\clearpage       
\phantomsection

\printbibliography

\end{document}


\begin{table}[ht]
\centering
\caption*{Supplementary Table S1.Summary of Other Target Frequencies, Periods, KZFT Settings, and Result}
\label{tab:kzft_vbpbb}

\resizebox{\textwidth}{!}{%
\begin{tabular}{ c c c c c c c }
\hline
\multicolumn{1}{c}{} & \multicolumn{2}{c}{KZFT Arguments} & 
\multicolumn{2}{c}{VBPBB 95\% CI Band} & 
\multicolumn{2}{c}{GSBB 95\% CI Band} \\ 
\cline{2-3} \cline{4-5} \cline{6-7}
$Target Frequency$  & $m$ & $k$ & Crest & Trough & Crest & Trough  \\
\hline
0.021904 & 733 & 2 & (-0.017,0.044) & (-0.044,0.017) & (-0.796,0.794) & (-0.861,0.571)\\
0.02738 & 731 & 2 & (-0.021,0.029) & (-0.029,0.021) & (-0.759,0.824) & (-0.853,0.552)\\
0.035594 & 1771 & 2 & (-0.016,0.033) & (-0.033,0.015) & (-0.163,0.824) & (-0.875,0.01)\\
0.038332 & 1461 & 2 & (-0.015,0.022) & (-0.022,0.015) & (-0.735,0.786) & (-0.866,0.604)\\
0.05476 & 731 & 2 & (-0.018,0.03) & (-0.03,0.018) & (-0.729,0.743) & (-0.82,0.626)\\
0.060236 & 731 & 2 & (-0.026,0.03) & (-0.03,0.027) & (-0.73,0.724) & (-0.823,0.584)\\
0.109521 & 1237 & 2 & (-0.027,0.029) & (-0.029,0.026) & (-0.758,0.719) & (-0.812,0.638)\\
0.112757 & 1237 & 2 & (-0.034,0.037) & (-0.037,0.035) & (-0.758,0.719) & (-0.812,0.638)\\
0.172993 & 367 & 2 & (-0.032,0.041) & (-0.041,0.034) & (-0.749,0.724) & (-0.813,0.567)\\
0.183945 & 487 & 2 & (-0.029,0.043) & (-0.043,0.03) & (-0.735,0.714) & (-0.817,0.603)\\
\hline
\end{tabular}
} 
\end{table}

\begin{figure}[ht] 
    \centering
    \includegraphics[width=0.7\textwidth]{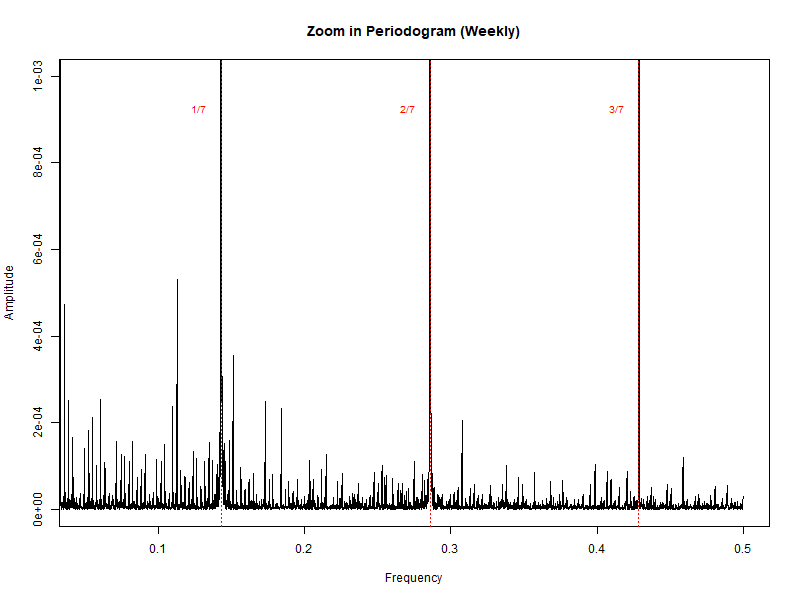}
    \caption*{Supplementary Figure S1. Zoomed-in periodogram of the IHD hospitalization rate time series, with the weekly frequency and its harmonics indicated.}
    \label{fig:Periodogram_Weekly}
\end{figure}

\begin{figure}[ht] 
    \centering
    \includegraphics[width=0.7\textwidth]{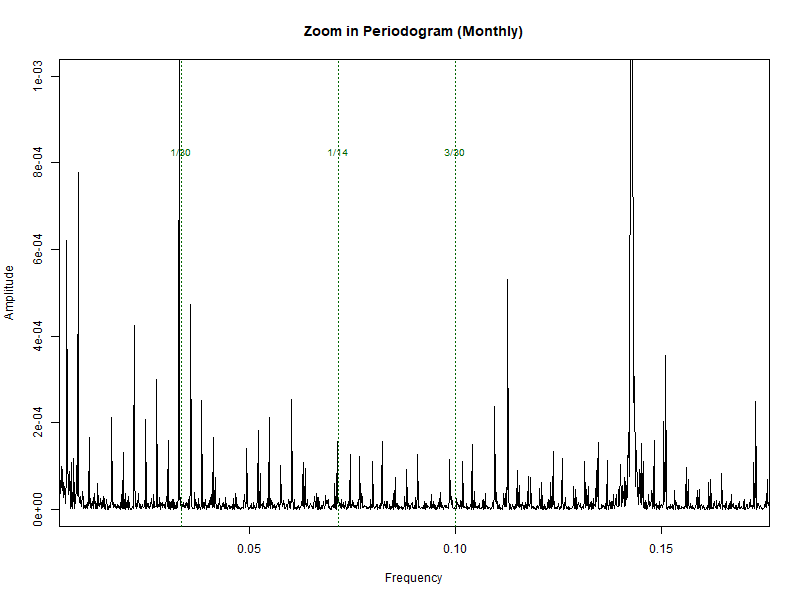}
    \caption*{Supplementary Figure S2. Zoomed-in periodogram of the IHD hospitalization rate time series, with the monthly frequency and its harmonics indicated.}
    \label{fig:Periodogram_monthly}
\end{figure}

\begin{figure}[ht] 
    \centering
    \includegraphics[width=0.7\textwidth]{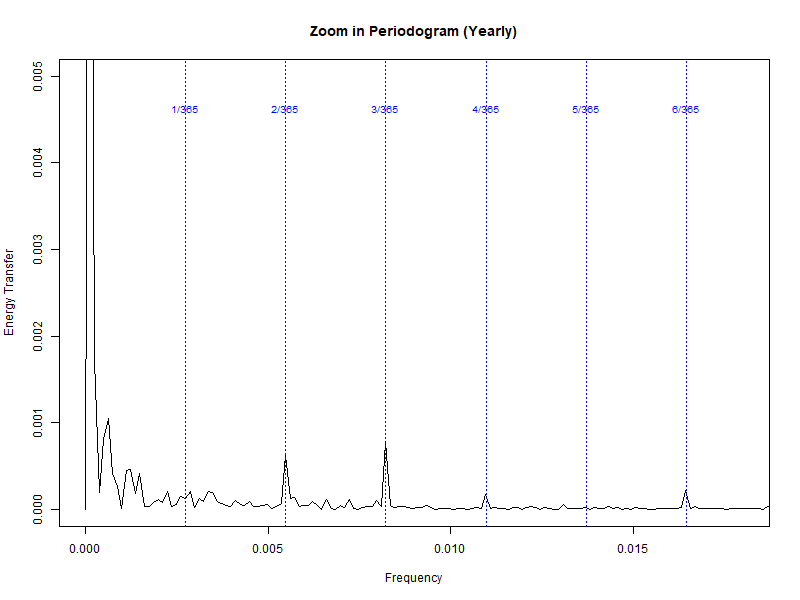}
    \caption*{Supplementary Figure S3. Zoomed-in periodogram of the IHD hospitalization rate time series, with the annual frequency and its harmonics indicated.}
    \label{fig:Periodogram_annual}
\end{figure}